# Pseudogap associated with precursor pairing

Tian De Cao[*]

*Department of physics, Nanjing University of Information Science & Technology, Nanjing 210044, China*

This work argues that the off-diagonal long range order (ODLRO) is not necessary for showing superconductivity while the electron pairing around Fermi surface is sufficient for superconductivity. It is shown that there exists the pseudogap state associated with the electron pairing in real space and the high temperature superconductivity could be only found in the metallic region near the Mott metal–insulator transition (MIT).

One of the open questions is whether the pseudogap is associated with the precursor pairing. There are some positive suggestions from experiments which include the specific-heat measurement [1], while there some negative indications which include the angle resolved photoemission spectroscopy (ARPES) [2]. The pseudogap is initially discovered in cuprate superconductors, and may be also discovered in iron-based superconductors [3]. The debate may go to peace when the precursor pairing is proved in theory. There exist two main theoretical scenarios for the explanation of pseudogaps in high temperature superconductors. The first is based upon the model of Cooper pairs formation already above the critical temperature of superconducting transition [4,5]. The second one assumes that the appearance of the pseudogap state is due to fluctuations of some short-range order [6,7]. To understand the precursor pairing and the pseudogap, we should discuss the pairing in real space, and the latter is related to the off-diagonal long range order (ODLRO) which is presented to determine whether superconductivity is included in a model[8,9]. Some works question whether the Hubbard model and the t-j model include superconductivity [10]. Yang's articles are associated with the relations between BEC and ODLRO, the ones between ODLRO and superconductivity, and the ones between BEC and superconductivity, while we think these relations have not been established exactly in theory. One will find that the BEC only describes the Bose-particle feature of Cooper pairs, while Cooper pairs are not equal to Bose particles; the ODLRO expresses the correlations between Cooper pairs, while Cooper pairs originate from the many-body effect. This work suggests possible theoretic evidence of precursor pairing, and argues

[*]Corresponding author.
[*]E-mail address: tdcao@nuist.edu.cn (T. D. Cao).
[*]Tel: 011+86-13851628895

that the ODLRO (about pairing) is not necessary for superconductivity.

Not to say the hardships of the BCS theory, the most important idea of the BCS theory is that the superconductivity originates from the electron pairing around Fermi surface, and this should be taken for an evidence to examine other superconducting theories.

As seen in literatures, the singlet pairing function in real space is

$$F(l,l+\delta,\sigma,\tau-\tau')=<T_\tau d_{l\bar{\sigma}}(\tau)d_{l+\delta\sigma}(\tau')> \tag{1}$$

and the pairing function in momentum space is

$$F(k,\sigma,\tau-\tau')=<T_\tau d_{\bar{k}\bar{\sigma}}(\tau)d_{k\sigma}(\tau')> \tag{2}$$

where the vector symbols of wave vectors are neglected. Based on the transforms on a lattice one can find the relations

$$F(l,l+\delta,\sigma,\tau-\tau')=\frac{1}{N}\sum_k e^{ik\cdot\delta} F(k,\sigma,\tau-\tau') \tag{3}$$

$$F(k,\sigma,\tau-\tau')= \sum_\delta e^{ik\cdot\delta}F(0,\delta,\sigma,\tau-\tau') \tag{4}$$

To arrive at Eq.(4), $F(l,l+\delta,\sigma,\tau-\tau')=F(0,\delta,\sigma,\tau-\tau')$ is used with the Eq.(3). It is easy to find that $F(k)$=0 if $F(0,\delta)$=0. Similarly, $F(0,\delta)$=0 if $F(k)$=0. Then, to judge whether superconductivity is included in a model, is $F(k)$ equal to $F(0,\delta)$? No, the reason will be found. Firstly, if $F(k)$ are real and positive and limited to near the Fermi surface, due to the symmetry of the Brillouin Zone, we find $F(0,\delta)$ can be either positive or negative with Eq. (3). Secondly, $F(k)$ could be zero for $F(0,\delta)\neq 0$ with Eq.(4). For example, if $F(0,\delta)\sim \sum_Q \cos Q\cdot\delta$, we can obtain $F(k) =\sum_\delta e^{ik\cdot\delta}F(0,\delta)$=0 for $k\sim k_F$ but $Q\neq k_F$. Non-superconducting state does not require $F(k)$=0 for every $k$, but requires $F(k)$=0 for $k\sim k_F$, thus more forms of $F(0,\delta)$ could meet this requirement. This arrives at the conclusion that the precursor pairing could occur in real space if superconductivity is because of the electron pairing around Fermi surface (PAFS).

Speaking about the occurring of superconductivity, it suddenly occurred to us the concepts about "phase coherence", "Bose condensation", "off-diagonal long rang order", and so on, but the relation between these concepts has not been strictly established in theory. The ODLRO is discussed below.

It is easy to find the differences between ODLRO and PAFS, but we should give an explanation. One could introduce various correlation functions. The so-called off-diagonal long rang order is proposed by Yang, and it is described by the pair-pair correlation function similar to

$$P(\delta, \tau - \tau') = <T_\tau d_{l\bar{\sigma}}(\tau) d_{l\sigma}(\tau) d^+_{l+\delta\sigma}(\tau') d^+_{l+\delta\bar{\sigma}}(\tau)> \qquad (5)$$

One often questions whether superconductivity was included in a model with the function $P(\delta)$ and obtains negative results, which may be a misleading.

Following Eq.(5), we will take the number of cell for an unit, $N=1$. It is hard to find the simple dependence of the pairing function $F(k)$ on the pair-pair correlation function $P(\delta)$ by complex calculations, thus we give a simple approximation

$$P(\delta, \sigma, \tau - \tau')$$
$$= \sum_{k_1,k_2} <T_\tau d_{\bar{k}_1\bar{\sigma}}(\tau) d_{k_1\sigma}(\tau)><T_\tau d^+_{k_2\sigma}(\tau') d^+_{\bar{k}_2\bar{\sigma}}(\tau')> + \sum_{k_1,k_2} e^{-i[k_1+k_2]\delta} <T_\tau d_{k_1\sigma}(\tau) d^+_{k_1\sigma}(\tau')><T d_{k_2\bar{\sigma}} d^+_{k_2\bar{\sigma}}(\tau')>$$

Consider $\tau - \tau' \to 0$ and neglect the off-diagonal part of Green's function, we find

$$P(\delta, \sigma, 0) = [\sum_{k_1} F(k_1, \sigma, 0)][\sum_{k_2} F^+(k_2, \sigma, 0)] + [\sum_{k_1} e^{-ik_1\delta} n_{k_1\sigma}][\sum_{k_2} e^{-ik_2\delta} n_{k_2\bar{\sigma}}] \qquad (6)$$

It looks like $\sum_k e^{-ik\delta} n_{k\sigma} \to 0$ when $\delta$ is large enough. If $F^+ = F$ =real constant number, it seems $F$ =0 as soon as $P(\delta, \sigma, 0)$ =0, and this seems a proving for one to take $P(\delta, \sigma, 0)$ as an evidence of superconductivity. However, generally speaking, $\sum_{k_1} e^{-ik_1\delta} n_{k_1\sigma} \neq 0$ and $F^+ \neq$ real constant number as shown in strong coupling theory and so on, thus one cannot obtain $F$ =0 from $P(\delta, \sigma, 0)$ =0 for any $\delta$. Contrary to someone's expectation, we obtain $F(k_F) \neq 0$ for $P(\delta, \sigma, 0)$ =0 even if $\delta$ is small enough in Eq. (6). In the same way, we can obtain $P(\delta, \sigma, 0) \neq 0$ for $F(k_F)$ =0, thus $F(k_F) \neq 0$ is different from $P(\delta, \sigma, 0) \neq 0$. Of course, we usually finds $F \neq 0$ from $P(\delta, \sigma, 0) \neq 0$ when $P(\delta, \sigma, 0)$ has some fluctuations in real space, thus it is not strange for $P(\delta, \sigma, 0) \neq 0$ to mean Meissner effect and flux quantization [11]. That is to say, $P(\delta, \sigma, 0) \neq 0$ usually favors superconductivity for a large $\delta$ while $P(\delta, \sigma, 0)$ =0 does not reject superconductivity. If $F(k_F) \neq 0$, we also find the order function $\psi(x) = |\psi(x)| e^{i\theta(x)}$, thus some appearances like the Meissner effect and the flux quantization can be explained. However, $F(k_F) \neq 0$ does not mean $P(\delta, \sigma, 0) \neq 0$ as discussed above.

The pseudogap is another open question. As discussed above, if the pairing function $F(k_F)=0$, the electron systems are not in the superconducting states. The following discussion will show that this case corresponds to a pseudogap state. As an example, we take the Hubbard model

$$H = \sum_{<l,\delta>\sigma} t_{l,l+\delta} d^+_{l\sigma} d_{l+\delta\sigma} + \frac{1}{2} U \sum_{l,\sigma} d^+_{l\sigma} d_{l\sigma} d^+_{l\bar{\sigma}} d_{l\bar{\sigma}} \qquad (7)$$

It can be rewritten in

$$H = \sum_{k,\sigma} \xi_k d^+_{k\sigma} d_{k\sigma} + \sum_q U\hat{\rho}(q)\hat{\rho}(-q) - \sum_q U\hat{S}(q)\hat{S}(-q) \qquad (8)$$

where the charge operator $\hat{\rho}(q) = \frac{1}{2}\sum_{k,\sigma} d^+_{k+q\sigma} d_{k\sigma}$ and the spin operator $\hat{S}(q) = \frac{1}{2}\sum_{k,\sigma} \sigma d^+_{k+q\sigma} d_{k\sigma}$ in the wave vector space when we denote $k \equiv \vec{k}$. Other Green's functions are defined in

$$G(k\sigma, \tau-\tau') = -<T_\tau d_{k\sigma}(\tau) d^+_{k\sigma}(\tau')> \qquad (9)$$

$$F^+(k\sigma, \tau-\tau') = <T_\tau d^+_{k\sigma}(\tau) d^+_{\bar{k}\bar{\sigma}}(\tau')> \qquad (10)$$

If we neglect the effect of correlations, we only obtain the zero result $F^+=0$. Considering the effects of correlations but following the approximations presented by Abrikosov et al [12], we must establish the dynamic equations of many-particle correlation functions such as $\partial_\tau <T_\tau \hat{S}(q) d_{k+q\sigma} d^+_{k\sigma}(\tau')>$ and $\partial_\tau <T_\tau \hat{\rho}(q) d_{k+q\sigma} d^+_{k\sigma}(\tau')>$. These calculations arrive at the equations

$$[-i\omega_n + \xi_k + \sum_q \frac{P(k,q,\sigma)}{i\omega_n - \xi_{k+q}}] G(k\sigma, i\omega_n)$$

$$= -1 + \frac{U<\hat{\rho}(0)>}{-i\omega_n + \xi_k} + \frac{1}{2}\sum_q \frac{\xi_{k+q}-\xi_k}{-i\omega_n + \xi_{k+q}} UF(k+q\sigma, \tau=0) F^+(\bar{k}\bar{\sigma}, i\omega_n) \qquad (11)$$

and

$$[-i\omega_n - \xi_k - \sum_q \frac{P(k,q,\sigma)}{i\omega_n + \xi_{k+q}}] F^+(k\sigma, i\omega_n) = \frac{1}{2}\sum_q \frac{\xi_{k+q}-\xi_k}{-i\omega_n - \xi_{k+q}} UF^+(k-q,\sigma,\tau=0) G(\bar{k}\bar{\sigma}, i\omega_n) \qquad (12)$$

where

$$P(k,q,\sigma) = U<\hat{S}(-q)\hat{S}(q)>U - 2\sigma U<\hat{\rho}(-q)\hat{S}(q)>U + U<\hat{\rho}(-q)\hat{\rho}(q)>U \qquad (13)$$

The function $P(k,q,\sigma)$ will exhibit effects of correlations and $P(k,q,\sigma) \neq P(k,q,\bar{\sigma})$. Here $<\hat{S}(-q)\hat{S}(q)> \equiv <T_\tau \hat{S}(-q,\tau)\hat{S}(q,\tau-0^-)>$, $<\hat{\rho}(-q)\hat{\rho}(q)>$ and $<\hat{\rho}(-q)\hat{S}(q)>$ are similar to this expression. On the basis of Eq.(11) and (12), we can obtain the function $G(k\sigma, i\omega_n)$ and find the energy gap for $F^+ \neq 0$. For simplification, we take $F^+ \rightarrow 0$ in the expressions of $F^+$ and $G$, this does not mean $F^+ = 0$, and we obtain

$$F^+(k\sigma, \tau = 0) = -\frac{1}{\beta} \sum_n [i\omega_n + \xi_k + \Sigma^{(+)}(k,\sigma,i\omega_n)]^{-1} \cdot \frac{\Delta_+^{(-)}(k,\sigma,i\omega_n)}{i\omega_n - \xi_{k\bar{\sigma}} - \Sigma^{(-)}(k,\bar{\sigma},i\omega_n)} (1 - \frac{U<\hat{\rho}(0)>}{-i\omega_n + \xi_k}) \quad (14)$$

where

$$\Sigma^{(\pm)}(k,\sigma,i\omega_n) = \sum_q \frac{P(k,q,\sigma)}{i\omega_n \pm \xi_{k+q}}$$

$$\Delta^{(\pm)}(k,\sigma,i\omega_n) = \frac{1}{2} \sum_q \frac{\xi_{k+q} - \xi_k}{-i\omega_n \pm \xi_{k+q}} UF(k+q\sigma, \tau = 0) \quad (15)$$

$$\Delta_+^{(\pm)}(k,\sigma,i\omega_n) = \frac{1}{2} \sum_q \frac{\xi_{k+q} - \xi_k}{-i\omega_n \pm \xi_{k+q}} UF^+(k+q\sigma, \tau = 0)$$

To obtain an evident solution, we assume the on-site interaction $U$ to be not too large. Because the function $F^+$ is dominated by the frequency region where $\text{Im}\Sigma^{(+)}(k,\sigma,\omega) = 0$, Eq.(14) leads to

$$F^+(k\sigma, \tau = 0) = \sum_q U \frac{\Gamma(k,q,E_{k\sigma,+}) - \Gamma(k,q,E_{k\bar{\sigma},-})}{E_{k\sigma,+} - E_{k\bar{\sigma},-}} F^+(k+q\sigma, \tau = 0) \quad (16)$$

Where

$$\Gamma(k,q,E_{k\sigma,i}) = n_F(E_{k\sigma,i}) z^{(i)}(E_{k\sigma,i}) (1 + \frac{U<\hat{\rho}(0)>}{E_{k\sigma,i} - \xi_k}) \frac{\xi_{k+q} - \xi_k}{\xi_{k+q} + E_{k\sigma,i}} \quad (17)$$

for $i = \pm$, $z^{(\pm)}(\omega) = [1 + \sum_q P(k,q,\sigma)/(\omega \pm \xi_{k+q})^2]^{-1}$, and $\omega = E_{k\sigma,\pm}$ expresses the real solutions of $\omega \pm \xi_k \pm \sum_q P(k,q,\sigma)/(\omega \pm \xi_{k+q}) = 0$. It can be found that there is the solution $F^+(k\sigma) \neq 0$. It may be $F^+(k\sigma) = 0$ for $k \sim k_F$ but $F^+(k\sigma) \neq 0$ for other cases, this corresponds to the pseudogap state. Particularly, when the model parameters lead to $E_{k\sigma,+} \approx E_{k\bar{\sigma},-} \neq 0$ at some points in the momentum space, a high pseudogap temperature $T^*$ appears in Eq.(15). When $E_{k\sigma,+} = E_{k\bar{\sigma},-} = 0$ (the Fermi energy is defined as the chemical potential although the former is different from the latter a little.), Eq.(15) gives this expression

$$F^+(k_F, \tau = 0) = \frac{U}{2k_B T_c} z^{(\pm)}(0) \frac{\xi_{k_F} - U<\hat{\rho}(0)>}{\xi_{k_F}} \sum_q \frac{\xi_{k_F} - \xi_{k_F+q}}{\xi_{k_F+q}} F^+(k_F + q, \tau = 0) \quad (18)$$

It is possible for Eq.(18) to give a non-zero superconductivity transition temperature $T_c$. It seems that the superconductivity transition temperature increases with the on-site interaction, $T_c \propto U$. However, the conditions $E_{k\sigma,+} = E_{k\bar{\sigma},-} = 0$ require the chemical potential to be at the inside of the energy band, the systems show the metallic features, thus the on-site interaction could not be too large. In other words, our calculations suggest that the high temperature superconductivity should be found in the metallic region near the Mott metal–insulator transition (MIT), and this is in agreement with experiments.

In summary, there are the precursor pairing in real space, and this pairing corresponds to a type of pseudogap state. However, this calculation does not reject other pseudogaps associated with other short-range orders. Particularly, the off-diagonal long range order usually favors superconductivity while its disappearance does not reject superconductivity. Therefore, to distinguish whether superconductivity is included in a model, one could say yes on the ODLRO, not say no. The electron pairing around Fermi surface should be the evidence of various superconductivities, since this can explain Meissner effect, flux quantization and so on as shown in BCS theory despite the BCS theory is inappropriate for the high temperature superconductivity. We also show that the high temperature superconductivity should be found in the metallic region near the Mott metal–insulator transition (MIT).

# REFERENCES

[1] H. H. Wen, G. Mu, H. Q. Luo, et al, Phys. Rev. Lett. **103**(2009)067002.

[2] T. Kondo, R. Khasanov, T. Takeuchi, et al, Nature **457**(2009)296.

[3] H. Y. Liu, X. W. Jia, W. T. Zhang, et al, Chinese Physics Letters **25**(2008)3761.

[4] V. B. Geshkenbein, L. B. Ioffe, A. I. Larkin, Phys. Rev. B**55**(1997)3173.

[5] V. Emery, S. A. Kivelson, O. Zachar, Phys. Rev. B**56**(1997)6120.

[6] J. Schmalian, D. Pines, B. Stojkovic, Phys.Rev.Lett. **80**(1998)3839.

[7] A. P. Kampf, J. R. Schrieffer, Phys. Rev. B**41**(1990)6399.

[8] C. N. Yang, Rev. Mod. Phys. **34**(1962)694.

[9] C. N. Yang, Phys.Rev.Lett.**63**(1989)2144.

[10] G. Su, Phys. Rev. Lett. **86**(2001)3690.

[11] H. T. Nieh, G. Su and B. H. Zhao, Phys. Rev. B **51**(1994)3760.

[12] G. D. Mahan, Many-particle physics, p.778 (Plenum Press, New York, 1990).